# Multiplexed Biosensing of Proteins and Virions with Disposable Plasmonic Assays

Stephanie Wallace,[1] Martin Kartau*,[1] Tarun Kakkar,[1] Chris Davis,[3] Agnieszka Szemiel,[3] Iliyana Samardzhieva,[1] Swetha Vijayakrishnan,[3] Sarah Cole,[3] Giuditta De Lorenzo,[3] Emmanuel Maillart,[4] Kevin Gautier,[1] Adrian J. Lapthorn,[1] Arvind Patel,[3] Nikolaj Gadegaard,[2] Malcolm Kadodwala,[1] Edward Hutchinson,[3] Affar S. Karimullah*.[1]

1. School of Chemistry, University of Glasgow, Joseph Black Building, University Avenue, G12 8QQ, Glasgow, UK
2. James Watt School of Engineering, University of Glasgow, Rankine Building, Oakfield Avenue, G12 8LT, Glasgow, UK
3. MRC-University of Glasgow Centre for Virus Research, 464 Bearsden Road, G61 1QH, Glasgow, UK
4. HORIBA France SAS, 14, Boulevard Thomas Gobert - Passage Jobin Yvon, CS 45002 – 91120, Palaiseau, France

Corresponding Authors : Affar.Karimullah@glasgow.ac.uk, Martin.Kartau@glasgow.ac.uk

## Abstract
Our growing ability to tailor healthcare to the needs of individuals has the potential to transform clinical treatment. However, the measurement of multiple biomarkers to inform clinical decisions requires rapid, effective, and affordable diagnostics. Chronic diseases, and rapidly evolving pathogens in a larger population have also escalated the need for improved diagnostic capabilities. Current chemical diagnostics are often performed in centralised facilities and are still dependent on multiple steps, molecular labelling, and detailed analysis causing the result turnaround time to be over hours and days. Rapid diagnostic kits based on lateral flow devices can return results quickly but are only capable of detecting a handful of pathogens or markers. Herein we present the use of disposable plasmonics with chiroptical nanostructures, as a platform for low-cost, label-free optical biosensing with multiplexing and without the need for flow systems often required in current optical biosensors. We showcase detection of SARS-CoV-2 in complex media as well as an assay for Norovirus and Zika virus as an early developmental milestone towards high-throughput, single-step, diagnostic kits for differential diagnosis of multiple respiratory viruses and any other emerging diagnostic needs. Diagnostics based on this platform which we term 'Disposable Plasmonics Assays', would be suitable for low-cost screening of multiple pathogens or biomarkers in a near point of care setting.

The use of biomarkers for precision medicine allows great advancements to aid in the improvement of human health as well as the reduction in health-care costs.[1,2] Yet often, time, costs and capability of current technology limits the applicability of precision medicine concepts. The COVID-19 pandemic has demonstrated a need to monitor our health more regularly.[3] While daily testing is not an immediate necessity, a routine approach may become the norm to maintain social healthcare standards. Polymerase chain reaction (PCR) diagnostic approaches are currently the gold standard of infectious disease diagnostics, but their cost and turnaround time make them impractical for use in large scale routine testing. PCR diagnostics are also susceptible to shortage of oligonucleotide reagents as seen in the SAR-CoV-2 pandemic.[4] It has also been argued that sensitivity should be secondary compared to test frequency for large scale population testing.[5] Modelling of the SARS-CoV-2 pandemic indicates that such a strategy is theoretically capable of reducing the reproduction 'R' number of an epidemic.[5,6]

Rapid and economical detection of some pathogen components and biomarkers has been achieved using lower sensitivity tests such as lateral flow devices (LFDs) and traditional immunoassays (e.g. ELISA). LFDs in particular have made a significant step towards readily available mobile testing and are arguably the most cost effective and simplest testing technique albeit without quantification.[3,7] Yet, when applied to more than a single disease, these methodologies are either not high-throughput or require multiple reagents and lack ease of use. These technological limitations are a bottle neck in our progress towards being able to test rapidly for multiple pathogens with high-throughput and low costs. [3,7,8] LFDs require multiple antibodies plus label/colour producing reagents which often suffers from reduced sensitivity and reliability.[3,7,9] Furthermore, the diffusion based paper flow methodology limits the ability to add additional tests in a small area due to interference of flow paths and/or interference between sequential detection sites if positioned within a common flow path.[10] Hence, multiple testing with LFDs either involve complicated manufacturing methods or lead to large dimensions increasing cost and reducing mobility. Consequently, LFDs are often limited to detecting a few biomarkers.[10]

Optical based biosensing was long heralded as the best route to label-free point-of-care (PoC) multiplexed diagnostics.[11] Using the overlap between analytical chemistry and optical sensing, plasmonic sensors can detect interactions between a monolayer of surface immobilised binders and their target biomarkers. Due to their label-free sensing capability, the only reagents required are a buffer and the antibody/binder. A variety of plasmonic based techniques have been implemented such as Surface Plasmon Resonance (SPR), Surface Enhanced Raman Spectroscopy (SERS), Localised Surface Plasmon Resonance (LSPR), and plasmonics based colorimetric assays.[9] SPR based biosensors have been the most successful with surface functionalisation techniques to enhance specificity between biomarkers and surface attached ligands and achieved diagnostics of diseases such as AIDS and Hepatitis.[12,13] They only require binder immobilisation and buffers as reagents and are not dependant on adding any nanoparticles or additional steps such as mixing or rinsing. Recently, commercially available SPR biosensors have achieved portability such as systems produced by Affinite and Plasmetrix making SPR more accessible for analytical science.[14,15] However, high reagent volume requirements and complexity still persist.[16] LSPR devices were supposed to mitigate these issues but have often suffered difficulties such as reduced sensitivity to refractive indices. Complex nanostructure design could potentially allow for high quality factors and improved sensing performances but are restricted due to high manufacturing costs of consumables and reproducibility problems. A large number of LSPR sensors are still based on nanoparticles in solutions or colloidal Au based films but have started seeing success with companies such as LamdaGen and Nicoya for the biosensor market.[16–20] In terms of PoC diagnostics commercially, to the best of our knowledge, only Genalyte has been able to successfully use a photonics sensor to provide label-free multiplexed diagnostics, albeit still using a relatively complex and expensive consumable based on split-ring resonators on Si substrates.[21]

Herein, we report on the use of injection moulded nanopatterned polycarbonate templates, with complex nanostructure geometries, for use in multiplexed biosensing of proteins and virions as a proof of concept for the development of a multiplexed low-cost diagnostics platform. These low-cost templated plasmonic substrates (TPS) are capable of large-scale multiplexing with high surface sensitivity allowing for the development of multi-pathogen diagnostic assays that we call 'Disposable Plasmonic Assays' (DPAs). We performed label-free biosensing without any flow setup or microfluidics demonstrating the potential of DPAs to be used as a simplified platform for PoC diagnostics. The disposable plasmonics concept has previously been used for chiral plasmonic sensing, a technique that uses chiral nanostructures with biostructural sensitivity to measure protein binding interactions. In this work, we use the chiral nanostructures for their sharp optical rotation dispersion (ORD) response with their high figure of merit (FOM) that enable better automated peak detection and performance for label-free measurement. The chiral nanostructures used have high refractive index sensitivity (~400 nm/RIU) and surface sensitivities as their fields decay significantly by ~25 nm above the sensor surface. Using a hyperspectral polarimetry imaging instrument, we were able to measure multiple nanostructure arrays in a single experiment and mitigated the need for microfluidics by simply pipetting materials onto the sample surface. Through immobilisation of different protein binders on the arrays, this system is capable of multiplexed label-free assays that are free of any flow systems and could therefore enable single step testing. The measurement performance of the sensor platform was evaluated first, followed by label-free detection of protein binding events. The potential multiplexing capabilities were also demonstrated by the specific detection of antibodies for the SARS-CoV-2 spike glycoprotein S1 (anti-S1) and streptavidin (anti-streptavidin) in a single experiment with sequential addition of the targets. Lastly the detection of SARS-CoV-2, Noro and Zika virus (ZIKV) was demonstrated using functionalised antibodies.

# Results and Discussion

### Optical Characterisation

The TPS were generated by Au coating of an injection moulded plastic template, Figure 1 (C). The Au film takes on the shape of the nanostructured indentations on the plastic surface producing a metafilm. The process has been used previously and provides a high-throughput (12 samples every 6s) manufacturing process with remarkable resolution (~20 nm linewidths) that is similar to manufacturing of Blu-ray discs.[22] Specifically, we use chiral shuriken shaped indentations as the plasmonic resonator units, Figure 1 (A-B), used in previous studies.[23–26] The TPS used here are specifically designed to work with our imaging polarimetry system that recognises 9 locations labelled A to I, Figure 1(D), for multiplexing purposes. Each location has 2 nanostructured arrays, one with left-handed (LH) nanostructures and one with right-handed (RH) nanostructures. We can use LH and RH resonance shifts either to evaluate differences for chiral plasmonic sensing or use individual resonance shifts of all 18 nanostructured arrays to gather values for our biosensing measurements. The entire measurement region is approximately a 3x3 mm square and each nanostructured array has an area of 0.09 $mm^2$. Solutions are added using a pipette through ports in a custom designed fluidic chamber well, and no flow systems are incorporated in the setup (Supplementary Figure S1). The imaging instrument is capable of measuring ORD and reflectivity over the visible spectrum using hyperspectral imaging and polarisation dependent filters in ~ 5 minutes. A MATLAB script automatically evaluates the ORD peaks from the measured spectra and provides peak positions and resonance shift values, (Supplementary Figure S1).

The TPS display bisignate ORD with 2 inflection points that we label Peak 1 and Peak 2, Figure 1 (E). They also display a 'W' shaped reflectivity arising from plasmonic induced reflectance.[26] The ORD can be used as both a means of looking at resonance shifts as well as providing biostructural sensitivity as achieved in previous studies.[22,23] However, biostructural sensitivity requires the surface immobilised biomolecules to be aligned and achieve near homogenous orientation over the surface. This leads to an anisotropic dielectric layer surrounding the chiral nanostructures instead of one that is isotropic and leads to a measurable asymmetry.[24,25,27] Such constraints on the immobilisation of the biomolecules can be difficult in most functionalisation strategies and this restricts the practicality of generating assays with multiple binders for multiple targets. It is also important to note that different pathogens have different physical properties. While some viruses (for example adenoviruses or picornaviruses) are transmitted within a rigid icosahedral 'capsid' assembly of proteins, others (for example influenza viruses and SARS-CoV-2) are enclosed in an envelope of lipid membrane with viral proteins on its surface. These 'enveloped virions' are typically variable in size and shape and are flexible enough to be physically deformed, leading to a lack of consistent anisotropy in the overall structure at the metal dielectric boundary.[28] Achieving an immobilised layer of virions that are all well aligned to provide an anisotropic dielectric layer is not universally applicable to all virions and proteins.

However, chiral nanostructures, beyond their biostructure sensing capabilities, show improved (FOM) owing to the increased complexity in resonance mechanism and show improved refractive index sensitivities. Such properties improve sensing of traditional refractive index changes.[29] Measurement of chiral ORD response is also less susceptible to signal losses and variations generated by absorptive molecules when measuring through the sample. This can be useful given the birefringent polycarbonate substrates coated with >100 nm thick Au, restricts transmission measurements. The sharper ORD peaks, such as those shown by the shurikens, improve automation and collection of data from the experiment. Hence, our disposable plasmonic assays continue to use chiral optical properties for sensing refractive index variations to perform biosensing. Therefore to detect binding events we measure ORD from the shurikens by measuring reflectivity for four Stokes parameters (details in Supplementary Information). Changes in the two ORD peaks are measured as resonance shifts termed $\Delta\lambda$ and the value S that represents spacing in wavelength values between the two peaks. $\Delta S$ is the change in the spacing in comparison to the initial measurement and has previously been used as an additional parameter to measure protein interactions at the surface.[24,27]

We characterised the sensing performance of the TPS with sucrose and salt solutions. The results (Supplementary Figure S1 and Figure S2) showed a sensitivity value of ~430 nm/RIU which is between the general sensitivities of SPR (>1000 nm/RIU) and LSPR sensors (~100 nm/RIU) and similar to those shown by nanohole films.[30–32] Simulations of the nanostructures (Supplementary Figure S3) also show that the electric field intensities reduce to <15% of the maximum at ~25 nm from the surface, indicating that the structures have lower decay lengths than SPR (>100nm) and similar to LSPR (~5-10nm) sensors, indicating high surface sensitivities similar to LSPR sensors.[17] The electromagnetic confinement shown by LSPR and our metafilm make them less susceptible to bulk changes due to effects like temperature changes or additional proteins expected in serum like samples. It also potentially provides increased sensitivity to small molecules at low concentrations.[16,33] Hence, the shuriken metafilms combine sensing benefits of traditional LSPR and SPR biosensors.

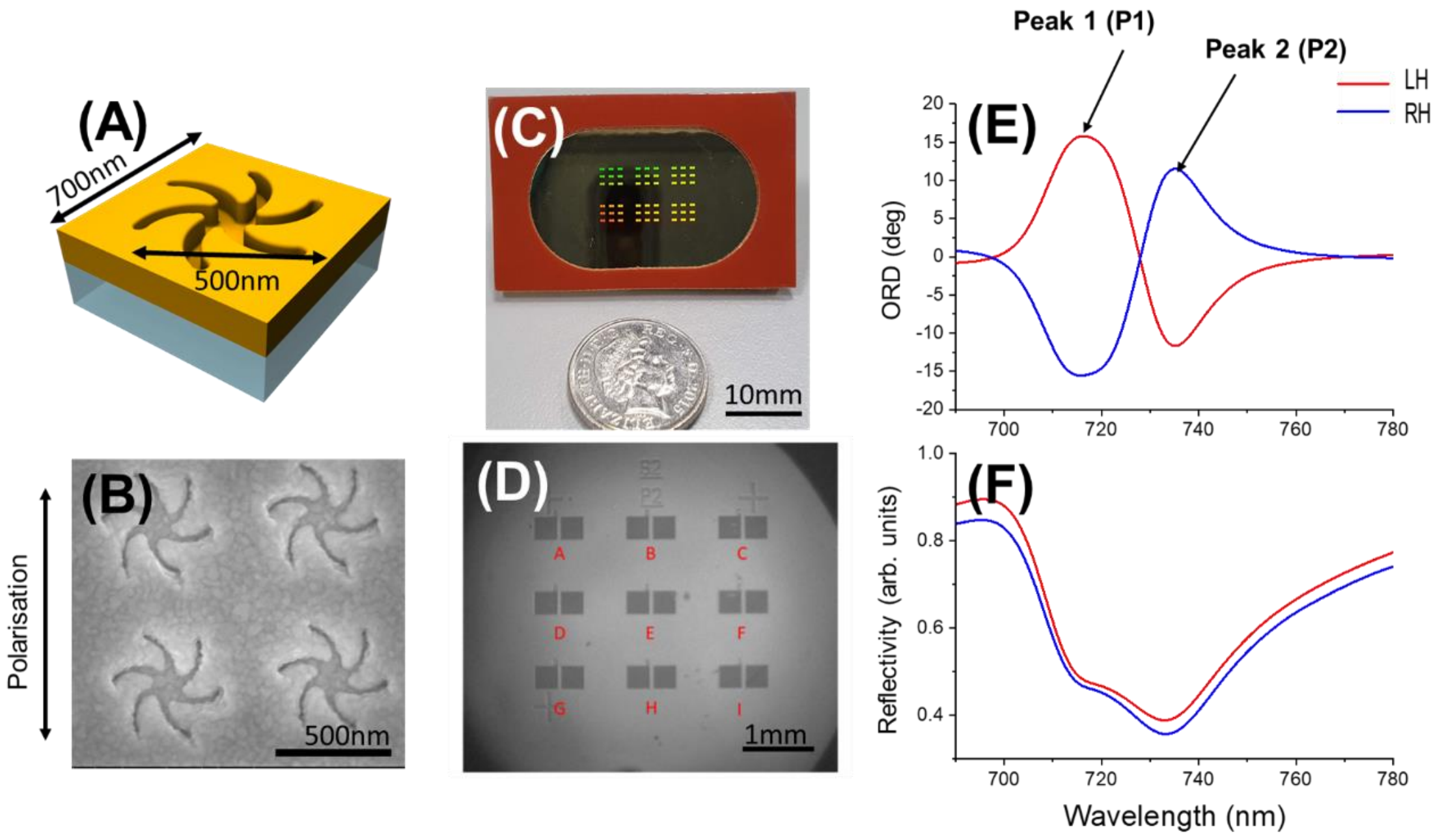

*Figure 1: (A) Top of the TPS film with shuriken indentations, as shown in (B) scanning electron microscopy images. (C) Sample with multiple experimentation locations compared to a 5p coin. (D) Arrays in a single experiment as viewed by the imaging instrument for multiplexing. (E) ORD (Peak 1 and Peak 2) and (F) reflectivity spectra produced by biosensor.*

## Biosensing

Protein-ligand interactions of streptavidin-biotin binding were measured to test the sensor. Streptavidin binds to biotin to form one of the strongest non-covalent interactions in nature.[34] As a tetramer, streptavidin has 4 binding sites for biotin. When binding to a biotinylated self-assembled monolayer (SAM), it is likely to bind to only 1 or 2 biotin sites on the surface at any time due to the symmetry of the streptavidin structure and due to the surface density of biotin moieties immobilised in the SAM.[35] Hence, a minimum of 2 vacant sites would be expected for additional biotin to bind to the protein. The streptavidin-biotin interaction therefore becomes an appropriate model system to study the performance of our sensor platform.

Biotinylated polyethylene glycol (PEG) thiols were immobilised with spacer molecules (methyl PEG thiol or MT-PEG) to create a SAM to functionalise the streptavidin onto the surface, Figure 2 (A). The spacer concentrations were optimised to completely inhibit non-specific interactions (Supplementary Figure S4). Streptavidin was added for 30 mins followed by a rinse with buffer and a single measurement. The peak values were evaluated by the software from the ORD spectrum during measurement and the $\Delta\lambda$ change from the first reference water measurements are used to plot the mean $\Delta\lambda$ values. Figure 2(B) shows the resonance shifts for each peak at each step of the experiment. The box plots (25-75% quartile shown by the box and max-min range by the whiskers for all 18 nanostructured arrays) show The SAM layer shows good adhesion to the surface with a mean $\Delta\lambda$ value of 1.6 nm for Peak 2 compared to buffer. The streptavidin (~55 kDa) is a medium sized protein and hence generates a mean of 3 nm $\Delta\lambda$ shift (Peak 2) compared to the SAM. An additional experiment performed for the same interaction using a new sample showed good repeatability (Supplementary Figure S5). Streptavidin immobilisation step was followed by the addition of biotin with an Atto-655 conjugate as outlined in Figure 2(A) (and Supplementary Figure S6).

Upon addition of biotin the mean resonance shifts negatively by 1.2 nm (Peak 2, 40% change in value) even though the biotin is bound to the surface immobilised tetrameric streptavidin as confirmed by the plasmonic enhanced fluorescence from the Atto-655 conjugated to the biotin, Figure 2 (C).[36] Given the extremely low dissociation constant of the streptavidin-biotin interaction, $10^{-15}$ M, it is highly unlikely for the biomolecules to be removed from the surface.[37] Focusing on the mean ΔS values, the relatively large streptavidin causes an increase of 0.4 nm. Yet the mean ΔS only reduced by 0.1 nm (25% change in value) for biotin binding to streptavidin. This change in ΔS is far less than the change shown by the mean Δλ, contradicting streptavidin dissociation from the surface. The repeat experiment with streptavidin conjugated with Alexa Fluor 647 (results in Supplementary Figure S5) showed comparable Δλ values. Hence, it can be assumed that the samples have similar amounts of streptavidin on the surface given the same protocols for the SAM were used. While the dyes are different, fluorescence images show the homogenous surface coverage. The results indicate that the streptavidin is still bound to the surface. We hence hypothesise that the negative Δλ values would potentially be the result of structural changes (compacting) in streptavidin upon binding to additional biotin.[38–40]

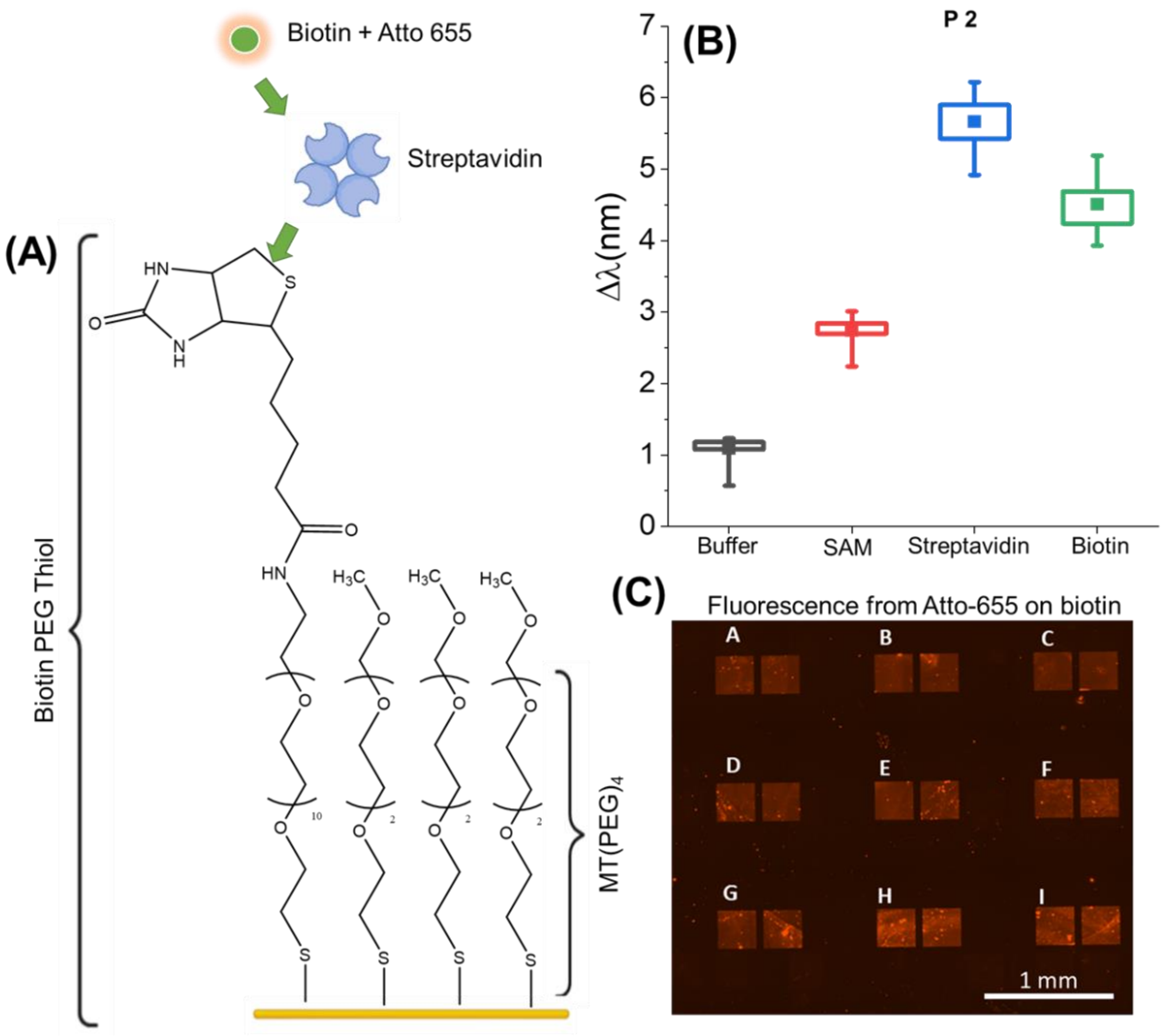

*Figure 2: (A) Functionalisation and experimental scheme: Biotin PEG Thiol/MT(PEG)$_4$ SAM functionalised to Au surface; binding of Streptavidin to functionalised biotin; binding of Atto-655 labelled biotin to streptavidin. (B) Results for Peak 2 resonance shifts for streptavidin binding to biotin PEG thiol SAM, followed by addition of biotin (Atto-655 conjugated). Initial measurements were taken in water then buffer (PBS). Hence, all biosensing measurements are taken relative to water. Peak 2 rinsed data only shown. (C) Fluorescence from the final Atto-655 conjugated biotin observed on the nanostructures. Fluorescence over the nanostructures is more prominent due to plasmonic enhancement of the fluorescence.*

## Multiplexing for Multi-Target Diagnostics

Multiplexing in relation to biomedical diagnostics can be defined as the simultaneous measurement of multiple analytes under the same set of conditions in a single experiment and sample.[16] A DPA antibody diagnostics proof-of-concept is generated by functionalising a single TPS in four separate regions (two each) by dropping 500 nL volumes of the specific histidine (His)-tagged proteins onto TPS coated with a SAM made using thiolated polyethylene glycol with a nitrilotriacetic acid end group (NTA-PEG-thiol) and a ethylene glycol thiol (EG-thiol) spacer with a 1:4 ratio. The NTA chelating agent can bind $Ni^{2+}$ which selectively binds the His-tagged proteins. As there are no separate compartments or fluidic systems for the binders in this DPA, the target antibodies were added to the chamber sequentially to evaluate specific target recognition by monitoring the locations of the individual binders, as shown in Figure 3 (D).

The first of the two protein binders used was streptavidin. The second protein binder was selected to show relevance to diagnostics related for SARS-CoV-2, the virus which causes COVID-19. The virus particle is covered with a large number of glycosylated spike (S) proteins, that form trimeric spikes. These are promising targets for COVID-19 antigen testing in the nasal mucus of infected individuals and its antibody tests are useful to evaluate post-infection as well. [41,42] We used spike 1 (S1) protein, a sub-unit of the overall spike protein (detail in Supplementary Information) as the binder for this purpose. Initial test of anti-S1 IgG antibody (Ab) targets binding to the S1 were performed with an artificially reconstituted mimic of human mucus. The artificial mucus which contains 0.2% mucin, 0.25 mg/ml haptoglobin and 0.50 mg/ml transferrin in phosphate buffer saline (PBS), showed non-specific binding with mean $\Delta\lambda$ increasing by 0.2 nm (Peak 2, from 2.6 nm to 2.8 nm). This was much smaller than the specific interaction with the target anti-S1 IgG (in serum), with mean $\Delta\lambda$ increasing by an additional 1.2 nm (to 4 nm), Figure 3 (B). As a reference, an additional experiment without serum was performed and is shown in Supplementary Figure S9.

After successful confirmation of the immobilisation strategy and testing S1-protein interaction with anti-S1 Ab in artificial mucus, the multiplexed DPA was completed with streptavidin immobilised on locations A and C, and S1-protein immobilised on locations G and I, Figure 3 (D). All other locations were ignored. Measurements after PBS rinsing are shown in Figure 3 (C) where all values are the average of the two locations for each target relative to the initial buffer measurements (complete data in Supplementary Figure S10-11 and Table S5). Each target Ab (1 µM) was spiked in artificial mucus and was introduced sequentially (Step 1 for anti-streptavidin and Step 2 for anti-S1) into the chamber and left for 15 minutes, then rinsed with PBS after which the measurements were performed. Addition of the anti-streptavidin Ab, Step 1, incurred minimal non-specific binding of this Ab to the S1 protein (change in mean $\Delta\lambda$ is 0.1 nm Peak 2), whilst specific binding to streptavidin showed a mean $\Delta\lambda$ change of 3 nm (Peak 2). Immobilisation of the Anti-S1 Ab thereafter bound only to the S1 protein giving a change in mean $\Delta\lambda$ of 1.2 nm, while the streptavidin regions showed change in mean $\Delta\lambda$ of only 0.3 nm due to potential non-specific interactions. These results validate the specific detection capabilities of this multiplexing setup with the potential to detect various biomolecules within one experimental setup using binders coated onto specific regions without the need of kinetic measurements or flow systems required in SPR.

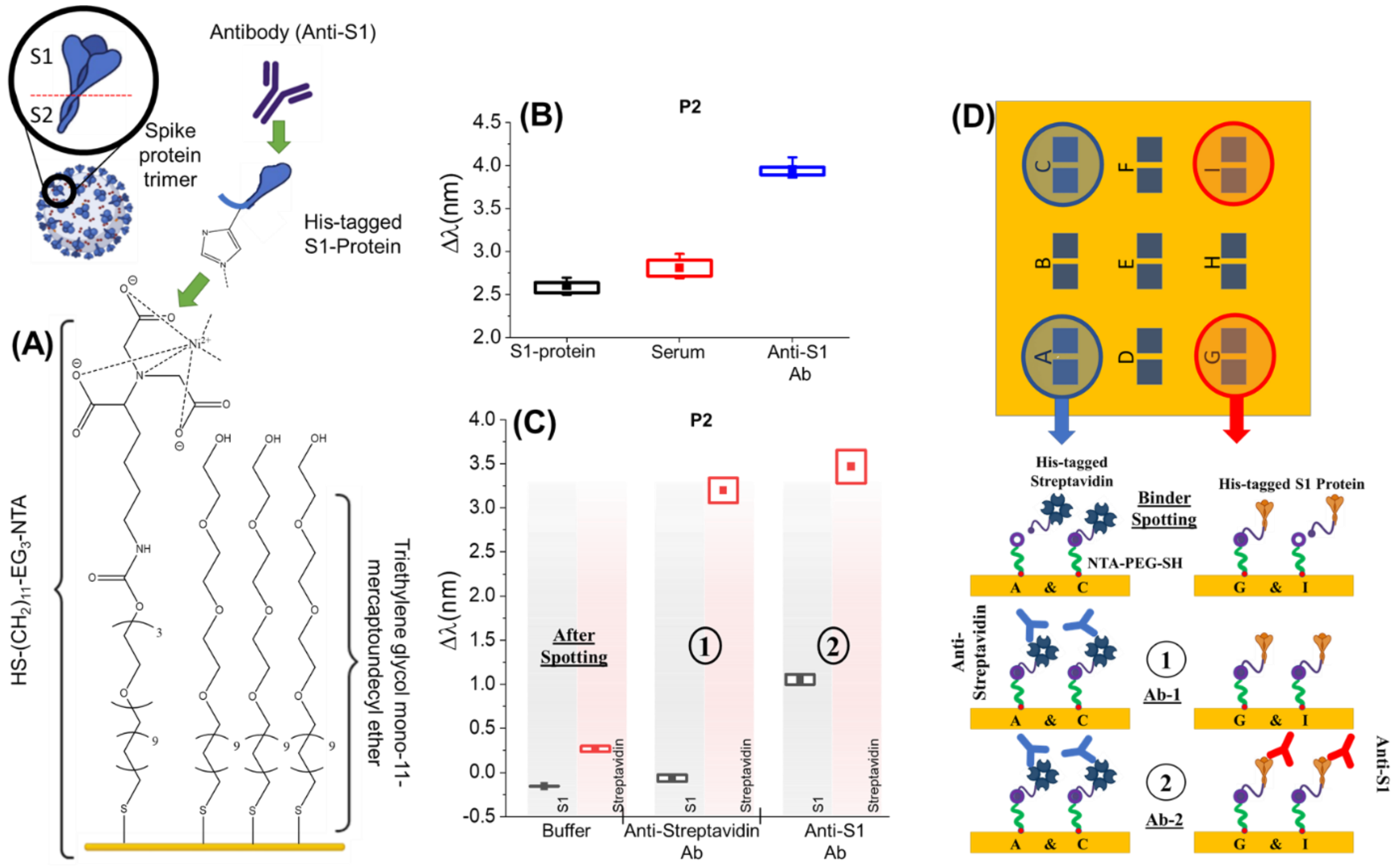

*Figure 3: (A) Functionalisation and experimental scheme of functionalised NTA/EG-thiol SAM being used to bind histagged S1-protein that is then used as an antigen test to detect anti-S1 IgG antibodies. (B) Results for 1μM S1-protein binding to pre-functionalised NTA/EG-thiol SAM, followed by addition of artificial mucus, which was then spiked with 1μM Anti-S1 Ab. Mean Δλ values were taken relative to water from all 18 nanostructure arrays. (C) Multiplexed DPA biosensing results showing mean Δλ values with sequential addition of target antibodies for histagged S1-protein (red regions) and histagged streptavidin (blue regions) spotted onto pre-functionalised NTA/EG-thiol SAM. This is followed by addition of spiked artificial mucus with Step 1) 1μM Anti-streptavidin Ab and then Step 2) 1μM Anti-S1 Ab. Peak 2 rinsed data only. Each box shows a data set of all 4 nanoarrays functionalised by a single binder (D) Graphical description of the experiment with sequential addition of the two targets.*

It should be noted from Figure 3 (C) that the resonance shifts exhibited for the Anti-S1 Ab are significantly smaller in comparison to those obtained for the Anti-streptavidin Ab, although both have a molecular weight of ~150 kDa.[43] This is likely due to the variation in coverage on the surface, hence, providing fewer epitopes for the antibodies to bind to. It is likely to be further compounded by variations in the antibody affinities for their targets.

## Detecting Virions

Next, we evaluated the biosensor platform for use in viral diagnostics. Additionally, we assess whether it could detect intact virions. These biological structures are substantially larger and more complex than individual proteins, and if they could be detected directly it would remove the need for lysis of samples, reducing the processing steps and reagents required. As an example, we targeted the SARS-CoV-2 virion for our DPA. Antibodies are the classic binders used in most immunoassay diagnostics and here we use the anti-S1 polyclonal antibody (pAb) as the binder. However, instead of functionalising antibodies through chemical moieties in our SAM, we use a simpler approach by immobilising Fab' antibody fragments directly onto the Au surface.[25] This reduces the need for additional functionalisation steps saving time and materials. SARS-CoV-2-binding antibodies were cleaved below the hinge region using immobilised pepsin. In the presence of Au, the F(ab')$_2$ fragments cleave to form F(ab') fragments and allow direct functionalisation onto the TPS as described in Figure 4(A) (further details in Supplementary Information).

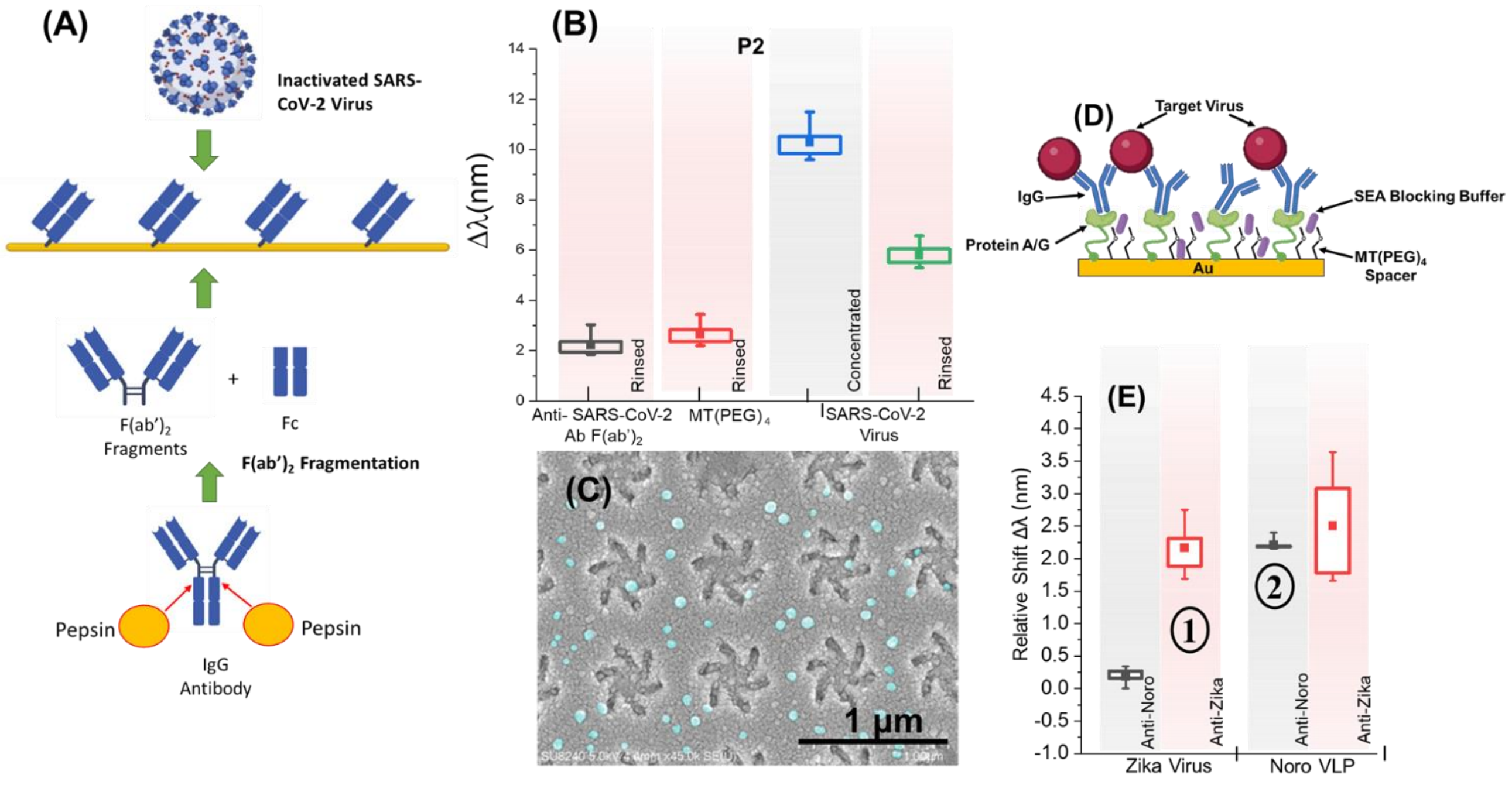

*Figure 4: (A) Description of polyclonal F(ab')$_2$ fragmentation and its functionalisation which is then used for detecting inactivated SARS-CoV-2 Virus. The MT(PEG)$_4$ spacer is not shown. (B) Results for inactivated SARS-CoV-2 virion detection experiment. The biosensing measurements were taken relative to an initial buffer measurement. Peak 2 data shown. (C) SEM of nanostructures with virion coverage across surface. Prominent viruses are highlighted in false colour (cyan). (D) Scheme for the multiplexed virus DPA using cysteine terminated protein A/G, sea blocking buffer and specific Abs for Norovirus and ZIKV capsid/envelop proteins. (E) Relative Δλ values when Zika virions are added in Step 1 and Norovirus VLPs are added in Step 2 to the sample. Each step shows specific binding of the targets to the respective regions where grey background represents regions with anti-Norovirus and red represents regions with anti-ZIKV. Each region includes 3 sites and hence 6 nanoarrays (Supplementary Figure S13).*

The F(ab')$_2$ fragment solution (Anti-SARS-CoV-2 spike Glycoprotein S1 Ab) was added to the TPA sample for 2 hours, followed by addition of 2mM MT(PEG)$_4$ for 2 hours to completely passivate the sample, preventing any non-specific binding of the virus which was added thereafter for 1 hour. The chamber was rinsed using buffer (PBS) and measurements were taken. Figure 4 (B) shows the mean Δλ values for Peak 2 (Peak 1 in Supplementary Figure S12). A 2.2 nm mean Δλ value was obtained for the binding of Anti-SARS-CoV-2 to the DPA, with a further 0.4 nm for the MT(PEG)$_4$ spacer.

The clinical isolate SARS-CoV-2-CVR-GLA-8 was amplified to a titre of $7.3x10^5$ plaque forming units (PFU) / ml (details in methodology). The virions were inactivated in formaldehyde for 30 min to allow safe handling, after which excess formaldehyde was quenched by diluting 1:1 with 50 mM Tris buffer resulting in final sample with ~$3x10^5$ inactivated virions per ml. After addition of the SARS-CoV-2 virus, a measurement with the viral solution was taken (concentrated) in addition to a measurement after a buffer rinse. Addition of the diluted SARS-CoV-2 solution to the DPA yielded a 7.7 nm Δλ when taking a measurement with the virus solution (concentrated) after 1 hour, which reduced by 3.8 nm upon rinsing with buffer. This reduction in resonance shift is indicative of the removal of non-specifically bound virions, implying that the final resonance shifts represent specifically bound virus material on the DPA. Scanning electron microscopy (SEM) images of the sample (Figure 4 (C)) show small spherical particles (~50-100 nm in diameter, false coloured in cyan) on the shuriken structures, indicative of virions, further reinforcing the resonance shifts obtained post rinsing.

Following the SARS-CoV-2 tests, multiplexed detection of virions was performed. A multiplexed virus DPA was developed for detection of Noro and ZIKV as an exemplar. The Ab functionalisation strategy implemented used a SAM incorporating protein A/G and MT(PEG)$_4$ spacer molecules to bind the Fc domain of IgG Abs as shown in Figure 4 (D) (further details in Supplementary Information). This strategy was found to improve mitigation of non-specific binding between the two virus targets compared to the Ab fragmentation protocol. Different monoclonal antibodies (mAb) were introduced onto specific regions of the DPA which were further optimised by the addition of SEA blocking buffer to mitigate non-specific binding. Two regions were coated specifically with a mouse anti-Norovirus mAb targeting the VP1 capsid protein and a mouse anti-ZIKV mAb targeting the 150-loop of the viral envelope protein (Figure S13 (B)). Figure 4 (E) shows the specific response from the regions where grey column represent results from the anti-Norovirus regions and red columns represent results from the anti-ZIKV regions. The $\Delta\lambda$ values are taken after a PBS rinse and are relative to when the experiment was started. In step 1, ZIKV particles ($4.8 \times 10^5$ PFU/ml) are introduced into the sample and in step 2, Norovirus virus like particles (VLPs) are introduced into the sample. Results show specific shifts for the two targets by their specific antibody coated regions. The anti-ZIKV region shows some non-specific interaction, however the mean $\Delta\lambda$ is much lower for the non-specific behaviour. The overall mean $\Delta\lambda$ for both targets are lower in comparison to the SARS-CoV-2 experiments. Inspection of the surface coverage by the virions and VLPs showed less coverage in comparison (Supplementary Figure S14) and suggests the need to further optimise of multiplexed virus DPAs in the future. Using $\Delta\lambda$ data from experiments measuring dilutions of the ZIKV (Supplementary Figure S15) we calculated the limit of detection (LOD) to be $3.1 \times 10^4$ PFU/ml for the ZIKV assay based on typical methods used for optical sensors.[44] These values are comparative to most optical biosensors and LFDs but still lower to modern more sensitive LFDs and electrochemical sensors based on newer methodologies.[45,46] Comparisons are further difficult as the method of defining the LOD is also highly variable and highly dependent on the target, their antibodies and assay methodologies. Our SARS-CoV-2 results showed much larger resonance shifts with higher virion coverage. The performance of each assay will hence depend on the Ab performance and surface coverage of the virions. Additionally, incorporating sandwich assay methodologies including usage of nanoparticles can improve the LOD where required but will increase the number of reagents. These results conclude the capability of DPAs to be used as multi-virus diagnostic tools.

# Conclusions

The concept of DPAs could provide a mass manufacturable route to cheaper and reliable label-free biosensors for diagnostics. Our shuriken metafilms have shown attributes of LSPR and SPR sensors and the biosensing results here prove the validity and accuracy of multiplexed plasmonic biosensing using these substrates. Our platform based on a hyperspectral imaging instrument and the TPS consumable was used successfully to detect protein-protein interactions with high sensitivity to interactions at the surface. Using this platform, a simple DPA was created with multiple functionalised binders for an antibody target detection exemplar that successfully detected antibodies in complex artificial mucus like conditions. The direct detection of virions, including multiplexed detection, without lysing or any additional labelling was also successfully demonstrated. Through direct functionalisation of the antibodies to the surface, we demonstrate further reduction of the number of reagents required to produce such assays. With increased multiplexing, DPAs can be developed for high-throughput target screening assays. Further development towards a more compact instrument will provide PoC capability with single-step drop testing DPAs for analytical work in laboratories or in-field diagnostics for multiple respiratory viruses, saving costs and improving healthcare. This work lays the foundation of this new technological platform that can provide a seamless transition from research to in-field application with a potential to alter the way modern diagnostics and precision medicine are practised.

# Methodology

### Fabrication of Templates

The TPS used are generated by first producing polycarbonate templates. These templates are created by first writing a pattern in PMMA using electron beam lithography (Raith) and then electroplating to generate a Ni shim. The shim is used as the master in the injection moulding to produce polycarbonate microscope slides with nano indented surfaces, that are the templates for the plasmonic metafilms. These templates are coated with a thin layer of gold (120 nm) through electron beam deposition (MEB-550s Plassys) to provide the metafilm that completes the TPS. Each TPS is then cleaned in a 25 Watt plasma oxygen asher for 30 seconds prior to any functionalisation. A customised fluidic well (GraceBio) with a glass cover slip is attached to the surface of the TPS for experimentation. Additional details of the fabrication can be found in previous work by Darling et al. [47,48]

### Optical Measurements

A custom-built microscope is used to image the TPS surfaces using a CMOS camera (FLIR) with a variable polariser, and a monochromatic light source (Spectral Photonics) polarised using a nanoparticle polariser (Thorlabs). More details in Supplementary Information. LabVIEW software is used to control the light source wavelength and capture data from 18 locations in the image, corresponding to the nanoarrays on the TPS, and generate the dispersion spectrums. The software calculates the peak positions, and a table is generated for all 18 locations to provide resonance peak wavelength values. Samples are placed on a stage with multi axis alignment features and alignment was performed to achieve even illumination and ORD with equal and opposite graphs from both left-handed and right-handed nanostructure arrays. For each experimental step, 3 measurements were taken (of all nanopatterned arrays) and the average was used to measure the resonance position for each location. Graphs for each step were produced using the mean and standard deviation for all 18 locations.

### Streptavidin Experiments

Solution for the self-assembled monolayer (SAM) functionalisation were prepared using a 1:4 ratio of Biotin(PEG) Thiol: MT(PEG)$_4$ Thiol (Polypure 41156-1095; ThermoFisher 26132) with the constituents having a total 100 mM concentration in phosphate buffer saline (PBS, ThermoFisher). The sample was incubated in this solution for 24 h, followed by rinsing with PBS and measurements with PBS, for the starting reference values. Streptavidin (ThermoFisher 21122), at 1 µM, was prepared in PBS and added to the sample for 2 h. Measurements were taken before and after PBS rinsing. 1 µM Atto-655-Biotin (Sigma-Aldrich 06966) was also prepared in PBS and added to the sample for a further 2 hr period. Again, measurements were taken pre- and post-rinsing using PBS.

## S1 Protein Experiments

The SAM was prepared using a 1:4 ratio of HS-$(CH_2)_{11}$-$EG_3$-NTA: $HS(CH_2)_{11}(OCH_2CH_2)_3OH$ (Prochimia TH007-002; Sigma-Aldrich 673110), with the constituents having a total concentration of 1 mM in 95% ethanol. Following 4-5 h incubation of the samples in this solution, they were rinsed with 95% ethanol and incubated for a further 5 minutes in 1 mM aqueous NaOH. The samples were then rinsed with water and incubated in 40 mM $NiSO_4$ for 1 hr. Finally, the samples were rinsed with HEPES buffered saline (HBS) and water and dried with nitrogen. The fluidic chamber was attached and an initial reference measurement was taken of the SAM following rinsing with PBS and Tween20 (surfactant rinsing solution). The 125 kDa recombinant human coronavirus SARS-CoV-2 Spike glycoprotein S1 (Abcam ab273068) was prepared at a concentration of 0.2 µM in PBS and applied to the sample for 1 hr. Measurements were taken with the anchored S1 on the surface pre- and post-rinsing. A solution of 0.2% (w/v) mucin from bovine submaxillary glands (Sigma-Aldrich M3895), 0.25 mg/ml haptoglobin (Merck Sigma-Aldrich #H3536) and 0.50 mg/ml transferrin (Merck Sigma-Aldrich T3309) artificial mucus was prepared in PBS and added to the sample for 15 minutes and biosensing measurements taken. 0.2 µM anti-SARS-CoV-2 spike glycoprotein S1 mAb (Abcam ab275759) in artificial mucus was applied to the sample for 1 hr. The protein-protein interaction was measured pre- and post-rinsing.

For the multiplexing setup, the SAM was prepared in a 1:4 ratio as before. A 1 µM solution of S1 protein (Abcam ab273068) in PBS was prepared, and 0.5 µL was spotted onto specific regions of the TPS. Recombinant His-tagged streptavidin, (Prospec Pro-621) was also prepared at 1 µM in PBS and spotted onto another two regions of the TPS. These solutions were left on the sample for 1 hr, followed by PBS and Tween20 rinsing. A fluidic chamber was then attached and measurements were performed. 1 µM anti-streptavidin antibody (Sigma-Aldrich S6390) was prepared in PBS and added to the sample for a 1 hr period, followed by addition of 1 µM anti-S1 antibody (Abcam ab275759) for a further 1 hr. Measurements were taken for both antibodies pre- and post-rinsing.

## Isolation of SARS-CoV-2 Virus from Clinical Sample

SARS-CoV-2-CVR-GLA-8 virus (the clinical isolate Gla8) was isolated from nasal swabs from SARS-CoV-2-infected individuals. The sample was isolated by co-author Chris Davis, from a patient with consent given to the ISARIC4C consortium (https://isaric4c.net/). The ethical approval for sample collection and isolation was given by the Scotland A Research Ethics Committee (reference 20/SS/0028). The samples were transported in viral transport medium (VTM) mixed 1:4 in Dulbecco's Modified Eagle Medium (DMEM) supplemented with 2% fetal calf serum (FCS), 1% Penicillin-Streptomycin and 250ng/ml Amphotericin B (ThermoFisher scientific, cat# 10566016, 10499044, 15140122 and 15290018, respectively). The mixture was clarified at 3000rpm for 10mins and then used to inoculate Vero E6 cells (African Green monkey kidney cell line, from Michelle Bouloy, Institute Pasteur, France) in a 6 well plate. Samples were harvested between 48 and 96 hr post infection, depending on the extent of cytopathic effect (CPE). Viral presence was determined using a NEB Luna Universal Probe One-Step RT-qPCR Kit (New England Biolabs, E3006) and 2019-nCoV CDC N1 primers and probes (IDT, 10006713) and infectious titres by plaque assay. The viral sequence and the purity of the primary isolate were assessed using metagenomic next generation sequencing. Briefly, RNA was extracted from culture supernatant using a standard hybrid Trizol-RNeasy protocol (Thermofisher scientific cat# 15596018). Library preparations were completed from cDNA using Kapa LTP Library Preparation Kit for Illumina Platforms (Kapa Biosystems, cat# KK8232).

The sequencing of the libraries was carried out on Illumina's NextSeq 550 System (Illumina, cat# SY-415-1002). The resulting viral stock was designated CVR-GLA-8 (Genbank accession ON911332).

A virus working stock of CVR-GLA-8 was grown on A549-ACE2-TMPRSS2 and Vero E6 cell line as described previously.[49] The cells were maintained in DMEM-Glutamax supplemented with 10% fetal calf serum (FCS; Gibco) and non-essential amino acids (NEAA; Gibco) at 37°C in 5%(v/v) CO2, humidified incubator. Infections were carried out with the SARS-CoV-2-CVR-GLA-8 in monolayers of the Vero E6 cells, in medium supplemented with 2% FCS, and incubated at 32°C for 7 days, after which medium containing infectious virus was harvested. To assess the infectious titre, A549_ACE2_TMPRSS2 or Vero E6 cells in 12-well plates were infected with 10-fold dilutions of virus samples. After 1 hour incubation at 37°C, 1 ml of overlay comprising MEM, 2% FCS, 0.6 % Avicel (Avicel microcrystalline cellulose, RC-591) was added per well and incubated at 37°C for 3 days. Cell monolayers were fixed with 8 % formaldehyde and plaques were visualized by staining with 0.1% Coomassie Brilliant Blue (BioRad cat# 1610406) in 45% methanol and 10% glacial acetic acid. CVR-GLA-8 stock titre on A549_ACE2_TMPRSS2 cells was $7.3 \times 10^5$ PFU/ml, and $5.3 \times 10^4$ PFU/ml on Vero E6 cells.

## SARS-CoV-2 Inactivated Virus Experiments

Anti-SARS-CoV-2 spike glycoprotein S1 pAb (Abcam ab275759) was cleaved using a Pierce $F(ab')_2$ Micro Preparation kit (Thermofisher 44688) following manufacturer instructions, with the estimated antibody fragmentation being between 50-70%. Concentration calculations assume 50% conversion. Following this preparation, the $F(ab')_2$ fragment solution was added for 1 hr, and a measurement was taken (concentrated). A 2mM $MT(PEG)_4$ spacer solution was prepared in PBS and was added for 1 hr. Measurements were then taken pre- and post-rinsing with PBS.

Virus was inactivated by addition of 0.2 ml of formaldehyde (Fisher Scientific cat# F/1501/PB17) to 1 ml of virus (final formaldehyde concentration 6% (v/v)). After 30 min incubation at room temperature inactivated virus solution was removed from the CL3. Inactivated virus was stored at -20°C until further use. The inactivated virus was diluted 1:1 with TRIS buffer to quench the formaldehyde prior to application. The final solution was added to the fluidic chamber for 1 hr, and measurements were taken prior to and after rinsing with PBS.

All live virus procedures were performed in a Biosafety level 3-laboratory at the MRC-University of Glasgow Centre for Virus Research (SAPO/223/2017/1a).

## Isolation of Zika Virus from Clinical Sample

Zika virus strain PE243, initially isolated from a clinical source, was propagated in Vero E6 cells and its infectious titre determined by plaque assay as described.[50]

## Isolation of a mouse monoclonal antibody to the ZIKV envelope glycoprotein

Balb/c mice were immunized with a synthetic peptide corresponding to the 150-loop (amino acids 144 to 166) of ZIKV envelope protein and a monoclonal antibody (mAb), named ZkE3, was isolated using standard hybridoma technology. The specificity of mAb ZkE3 to the viral envelope 150-loop was confirmed by Western blot and ELISA (data not shown). A detailed characterisation of this mAb will be presented elsewhere. mAb ZkE3 was purified using protein G affinity chromatography for use in experiments described.

## Multiplexed Virus Experiments

The SAM was prepared using a 1:30 ratio of Protein A/G Cys-tagged recombinant protein (Prospec pro-1928, concentration of 6.3 µM) and $MT(PEG)_4$ spacer (189 µM, Thermofisher 26132) in ultrapure water. Following an incubation time of 16 hrs, the sample was rinsed with PBS. Mouse anti-Norovirus GI antibody (NativeAntigen MAB12495-100) at 1 µM, and anti-ZIKV virus antibody mAb ZkE3 at 1 µM, were prepared in PBS and added to separate regions of the sample for 2 hrs using an culture well inserts (Ibidi) to isolate the regions.  The sample was rinsed with PBS, and SEA blocking buffer (Thermo 37527) added to the sample for 30 min. The sample was rinsed with PBS, a fluidic chamber attached, and measurements performed. Medium containing 1.6 x$10^6$ PFU/ml ZIKV was diluted 70:30 (4.8 x $10^5$ PFU/ml) added to the sample for 1 hr. The sample was rinsed with PBS and measurements performed. Norovirus VLP solution (NativeAntigen REC31722-100) was added to the sample for 1 hr. The sample was rinsed with PBS and measurements performed.

## Acknowledgements

The authors would like to acknowledge the support by: EPSRC through grants EP/S001514/1, EP/S029168/1 and through the QuantIC funding scheme; MRC through a fellowship (MR/N008618/1 and MR/V035789/1), MRC grants (MC_PC_19026 and MC_PC_21023) and the Wellcome Trust Early Concepts in Development Scheme (219390/Z/19/Z). The authors acknowledge that parts of the experimental work was carried out by CRUSH: COVID-19 Antiviral Drug Screening and Resistance Hub at the MRC-University of Glasgow Centre for Virus Research (CVR).  Funding for this research was supported by a LifeArc COVID-19 award. The ZIKV-related work was funded by the Department of Health and Social Care using UK Aid funding and is managed by the NIHR (G.D.L. and A.H.P.) and by the UK Medical Research Council grant MC_UU12014/2 (A.H.P.). We would like to thank Ania Owsianka for help in generating MAb ZkE3. We would also like to acknowledge support by industrial collaborators Avacta Lifesciences, Horiba Lifesciences (Paris) and Pinpoint Medical.

## Supporting Information Available:

The following files are available free of charge.

File name: Supporting Information – Multiplexed Biosensing of Proteins and Virions with Disposable Plasmonic Assays. Contents include: Microfluidic well setup and instrumentation; Simulation for maximum electric field intensities; Additional biosensing data; Additional data on multiplexing experiment for multi target diagnostics; Additional data on multiplexing experiment for multi target diagnostics; Using antibodies to detect inactivated SARS-CoV-2 virus with plasmonic sensors; Additional information on multiplexed DPA for virus detection; Limit of detection (LoD).

## References

(1) Ginsburg, G. S.; Phillips, K. A. Precision Medicine: From Science To Value. *Health Aff* **2018**, *37* (5), 694–701. https://doi.org/10.1377/hlthaff.2017.1624.

(2) Slikker, W. Biomarkers and Their Impact on Precision Medicine. *Exp Biol Med* **2018**, *243* (3), 211–212. https://doi.org/10.1177/1535370217733426.

(3) Guglielmi, G. Fast Coronavirus Tests: What They Can and Can't Do. *Nature* **2020**, *585* (7826), 496–498. https://doi.org/10.1038/d41586-020-02661-2.

(4) Mercer, T. R.; Salit, M. Testing at Scale during the COVID-19 Pandemic. *Nat Rev Genet* **2021**, *22* (7), 415–426. https://doi.org/10.1038/s41576-021-00360-w.

(5) Larremore, D. B.; Wilder, B.; Lester, E.; Shehata, S.; Burke, J. M.; Hay, J. A.; Tambe, M.; Mina, M. J.; Parker, R. Test Sensitivity Is Secondary to Frequency and Turnaround Time for COVID-19 Screening. *Sci Adv* **2021**, *7* (1), 2020.06.22.20136309. https://doi.org/10.1126/sciadv.abd5393.

(6) Taipale, J.; Kontoyiannis, I.; Linnarsson, S. Population-Scale Testing Can Suppress the Spread of Infectious Disease. *medRxiv* **2021**, 2020.04.27.20078329. https://doi.org/10.1101/2020.04.27.20078329.

(7) Koczula, K. M.; Gallotta, A. Lateral Flow Assays. *Essays Biochem* **2016**, *60* (1), 111–120. https://doi.org/10.1042/EBC20150012.

(8) Zhang, Y.; Liu, X.; Wang, L.; Yang, H.; Zhang, X.; Zhu, C.; Wang, W.; Yan, L.; Li, B. Improvement in Detection Limit for Lateral Flow Assay of Biomacromolecules by Test-Zone Pre-Enrichment. *Sci Rep* **2020**, *10* (1). https://doi.org/10.1038/s41598-020-66456-1.

(9) Li, Z.; Leustean, L.; Inci, F.; Zheng, M.; Demirci, U.; Wang, S. Plasmonic-Based Platforms for Diagnosis of Infectious Diseases at the Point-of-Care. *Biotechnol Adv* **2019**, *37* (8), 107440. https://doi.org/10.1016/j.biotechadv.2019.107440.

(10) He, P.; Katis, I.; Eason, R.; Sones, C. Rapid Multiplexed Detection on Lateral-Flow Devices Using a Laser Direct-Write Technique. *Biosensors (Basel)* **2018**, *8* (4), 97. https://doi.org/10.3390/bios8040097.

(11) Anker, J. N.; Hall, W. P.; Lyandres, O.; Shah, N. C.; Zhao, J.; van Duyne, R. P. Biosensing with Plasmonic Nanosensors. *Nanoscience and Technology: A Collection of Reviews from Nature Journals* **2009**, 308–319. https://doi.org/10.1142/9789814287005_0032.

(12) Riedel, T.; Surman, F.; Hageneder, S.; Pop-Georgievski, O.; Noehammer, C.; Hofner, M.; Brynda, E.; Rodriguez-Emmenegger, C.; Dostálek, J. Hepatitis B Plasmonic Biosensor for the Analysis of Clinical Serum Samples. *Biosens Bioelectron* **2016**, *85*, 272–279. https://doi.org/10.1016/j.bios.2016.05.014.

(13) Diao, W.; Tang, M.; Ding, S.; Li, X.; Cheng, W.; Mo, F.; Yan, X.; Ma, H.; Yan, Y. Highly Sensitive Surface Plasmon Resonance Biosensor for the Detection of HIV-Related DNA Based on Dynamic and Structural DNA Nanodevices. *Biosens Bioelectron* **2018**, *100*, 228–234. https://doi.org/10.1016/j.bios.2017.08.042.

(14) Breault-Turcot, J.; Poirier-Richard, H. P.; Couture, M.; Pelechacz, D.; Masson, J. F. Single Chip SPR and Fluorescent ELISA Assay of Prostate Specific Antigen. *Lab Chip* **2015**, *15* (23). https://doi.org/10.1039/c5lc01045d.

(15) Choudhary, S.; Altintas, Z. Development of a Point-of-Care SPR Sensor for the Diagnosis of Acute Myocardial Infarction. *Biosensors (Basel)* **2023**, *13* (2). https://doi.org/10.3390/bios13020229.

(16) Lopez, G. A.; Estevez, M. C.; Soler, M.; Lechuga, L. M. Recent Advances in Nanoplasmonic Biosensors: Applications and Lab-on-a-Chip Integration. *Nanophotonics*. Walter de Gruyter GmbH January 1, 2017, pp 123–136. https://doi.org/10.1515/nanoph-2016-0101.

(17) Peixoto de Almeida, M.; Pereira, E.; Baptista, P.; Gomes, I.; Figueiredo, S.; Soares, L.; Franco, R. Gold Nanoparticles as (Bio)Chemical Sensors. *Comprehensive Analytical Chemistry* **2014**, *66*, 529–567. https://doi.org/10.1016/B978-0-444-63285-2.00013-4.

(18) Jarockyte, G.; Karabanovas, V.; Rotomskis, R.; Mobasheri, A. Multiplexed Nanobiosensors: Current Trends in Early Diagnostics. *Sensors (Basel)* **2020**, *20* (23), 1–23. https://doi.org/10.3390/S20236890.

(19) Mayer, K. M.; Hafner, J. H. Localized Surface Plasmon Resonance Sensors. *Chemical Reviews*. American Chemical Society June 8, 2011, pp 3828–3857. https://doi.org/10.1021/cr100313v.

(20) Gerion, D. Characterization and Performance of Commercial Localized Surface Plasmon Resonance Chips. *Plasmonics in Biology and Medicine IX* **2012**, *8234*, 823412. https://doi.org/10.1117/12.917241.

(21) Song, Y.; Huang, Y. Y.; Liu, X.; Zhang, X.; Ferrari, M.; Qin, L. Point-of-Care Technologies for Molecular Diagnostics Using a Drop of Blood. *Trends Biotechnol* **2014**, *32* (3), 132–139. https://doi.org/10.1016/J.TIBTECH.2014.01.003.

(22) Karimullah, A. S.; Jack, C.; Tullius, R.; Rotello, V. M.; Cooke, G.; Gadegaard, N.; Barron, L. D.; Kadodwala, M. Disposable Plasmonics: Plastic Templated Plasmonic Metamaterials with Tunable Chirality. *Advanced Materials* **2015**, *27* (37), 5610–5616. https://doi.org/10.1002/adma.201501816.

(23) Tullius, R.; Karimullah, A. S.; Rodier, M.; Fitzpatrick, B.; Gadegaard, N.; Barron, L. D.; Rotello, V. M.; Cooke, G.; Lapthorn, A.; Kadodwala, M. "Superchiral" Spectroscopy: Detection of Protein Higher Order Hierarchical Structure with Chiral Plasmonic Nanostructures. *J Am Chem Soc* **2015**, *137* (26), 8380–8383. https://doi.org/10.1021/jacs.5b04806.

(24) Kelly, C.; Tullius, R.; Lapthorn, A. J.; Gadegaard, N.; Cooke, G.; Barron, L. D.; Karimullah, A. S.; Rotello, V. M.; Kadodwala, M. Chiral Plasmonic Fields Probe Structural Order of Biointerfaces. *J Am Chem Soc* **2018**, *140* (27), 8509–8517. https://doi.org/10.1021/jacs.8b03634.

(25) Kakkar, T.; Keijzer, C.; Rodier, M.; Bukharova, T.; Taliansky, M.; Love, A. J.; Milner, J. J.; Karimullah, A. S.; Barron, L. D.; Gadegaard, N.; Lapthorn, A. J.; Kadodwala, M. Superchiral near Fields Detect Virus Structure. *Light Sci Appl* **2020**, *9* (1), 2047–7538. https://doi.org/10.1038/s41377-020-00433-1.

(26) Tullius, R.; Platt, G. W.; Khosravi Khorashad, L.; Gadegaard, N.; Lapthorn, A. J.; Rotello, V. M.; Cooke, G.; Barron, L. D.; Govorov, A. O.; Karimullah, A. S.; Kadodwala, M. Superchiral Plasmonic Phase Sensitivity for Fingerprinting of Protein Interface Structure. *ACS Nano* **2017**, *11* (12). https://doi.org/10.1021/acsnano.7b04698.

(27) Kelly, C.; Khosravi Khorashad, L.; Gadegaard, N.; Barron, L. D.; Govorov, A. O.; Karimullah, A. S.; Kadodwala, M. Controlling Metamaterial Transparency with Superchiral Fields. *ACS Photonics* **2017**, *5* (2), 535–543. https://doi.org/10.1021/acsphotonics.7b01071.

(28) Dadonaite, B.; Vijayakrishnan, S.; Fodor, E.; Bhella, D.; Hutchinson, E. C. Filamentous Influenza Viruses. *Journal of General Virology*. Microbiology Society August 1, 2016, pp 1755–1764. https://doi.org/10.1099/jgv.0.000535.

(29) Jeong, H.-H.; Mark, A. G.; Alarcón-Correa, M.; Kim, I.; Oswald, P.; Lee, T.-C.; Fischer, P. Dispersion and Shape Engineered Plasmonic Nanosensors. *Nat Commun* **2016**, *7*, 11331. https://doi.org/10.1038/ncomms11331.

(30) Chung, T.; Lee, S.-Y.; Song, E. Y.; Chun, H.; Lee, B. Plasmonic Nanostructures for Nano-Scale Bio-Sensing. *Sensors* **2011**, *11* (11), 10907–10929. https://doi.org/10.3390/s111110907.

(31) Guo, L.; Jackman, J. A.; Yang, H.-H.; Chen, P.; Cho, N.-J.; Kim, D.-H. Strategies for Enhancing the Sensitivity of Plasmonic Nanosensors. *Nano Today* **2015**, *10* (2), 213–239. https://doi.org/10.1016/j.nantod.2015.02.007.

(32) Yanik, A. A.; Cetin, A. E.; Huang, M.; Artar, A.; Mousavi, S. H.; Khanikaev, A.; Connor, J. H.; Shvets, G.; Altug, H. Seeing Protein Monolayers with Naked Eye through Plasmonic Fano Resonances. *Proceedings of the National Academy of Sciences* **2011**, *108* (29), 11784–11789. https://doi.org/10.1073/pnas.1101910108.

(33) Luchansky, M. S.; Washburn, A. L.; Martin, T. A.; Iqbal, M.; Gunn, L. C.; Bailey, R. C. Characterization of the Evanescent Field Profile and Bound Mass Sensitivity of a Label-Free Silicon Photonic Microring Resonator Biosensing Platform. *Biosens Bioelectron* **2010**, *26* (4), 1283–1291. https://doi.org/10.1016/J.BIOS.2010.07.010.

(34) Chivers, C. E.; Koner, A. L.; Lowe, E. D.; Howarth, M. How the Biotin–Streptavidin Interaction Was Made Even Stronger: Investigation via Crystallography and a Chimaeric Tetramer. *Biochemical Journal* **2011**, *435* (1), 55–63. https://doi.org/10.1042/BJ20101593.

(35) Ahlers, M.; Blankenburg, R.; Grainger, D. W.; Meller, P.; Ringsdorf, H.; Salesse, C. Specific Recognition and Formation of Two- Dimensional Streptavidin Domains in Monolayers: Applications to Molecular Devices. *Thin Solid Films* **1989**, *180* (1–2), 93–99. https://doi.org/10.1016/0040-6090(89)90059-X.

(36) Dong, J.; Zhang, Z.; Zheng, H.; Sun, M. Recent Progress on Plasmon-Enhanced Fluorescence. *Nanophotonics* **2015**, *4* (4), 472–490. https://doi.org/10.1515/nanoph-2015-0028.

(37) Piran, U.; Riordan, W. J. Dissociation Rate Constant of the Biotin-Streptavidin Complex. *J Immunol Methods* **1990**, *133* (1), 141–143. https://doi.org/10.1016/0022-1759(90)90328-S.

(38) Weber, P. C.; Ohlendorf, D. H.; Wendoloski, J. J.; Salemme, F. R. Structural Origins of High-Affinity Biotin Binding to Streptavidin. *Science (1979)* **1989**, *243* (4887), 85–88. https://doi.org/10.1126/science.2911722.

(39) González, M.; Bagatolli, L. A.; Echabe, I.; Arrondo, J. L. R.; Argaraña, C. E.; Cantor, C. R.; Fidelio, G. D. Interaction of Biotin with Streptavidin. *Journal of Biological Chemistry* **1997**, *272* (17), 11288–11294. https://doi.org/10.1074/jbc.272.17.11288.

(40) le Trong, I.; Wang, Z.; Hyre, D. E.; Lybrand, T. P.; Stayton, P. S.; Stenkamp, R. E. Streptavidin and Its Biotin Complex at Atomic Resolution. *Acta Crystallogr D*

*Biol Crystallogr* **2011**, *67* (9), 813–821. https://doi.org/10.1107/S0907444911027806.

(41) Huang, Y.; Yang, C.; Xu, X.; Xu, W.; Liu, S. Structural and Functional Properties of SARS-CoV-2 Spike Protein: Potential Antivirus Drug Development for COVID-19. *Acta Pharmacol Sin* **2020**, *41* (9), 1141–1149. https://doi.org/10.1038/s41401-020-0485-4.

(42) Mahmood, Z.; Alrefai, H.; Hetta, H. F.; A. Kader, H.; Munawar, N.; Abdul Rahman, S.; Elshaer, S.; Batiha, G. E.-S.; Muhammad, K. Investigating Virological, Immunological, and Pathological Avenues to Identify Potential Targets for Developing COVID-19 Treatment and Prevention Strategies. *Vaccines (Basel)* **2020**, *8* (3), 443. https://doi.org/10.3390/vaccines8030443.

(43) Janeway, C. A.; Traver, P.; Walport, M. The Structure of a Typical Antibody Molecule. In *Immunobiology: The Immune System in Health and Disease*; Garland Science, 2001.

(44) Cennamo, N.; Pasquardini, L.; Arcadio, F.; Lunelli, L.; Vanzetti, L.; Carafa, V.; Altucci, L.; Zeni, L. SARS-CoV-2 Spike Protein Detection through a Plasmonic D-Shaped Plastic Optical Fiber Aptasensor. *Talanta* **2021**, *233*. https://doi.org/10.1016/j.talanta.2021.122532.

(45) Xu, M.; Li, Y.; Lin, C.; Peng, Y.; Zhao, S.; Yang, X.; Yang, Y. Recent Advances of Representative Optical Biosensors for Rapid and Sensitive Diagnostics of SARS-CoV-2. *Biosensors*. 2022. https://doi.org/10.3390/bios12100862.

(46) Peto, T.; Affron, D.; Afrough, B.; Agasu, A.; Ainsworth, M.; Allanson, A.; Allen, K.; Allen, C.; Archer, L.; Ashbridge, N.; Aurfan, I.; Avery, M.; Badenoch, E.; Bagga, P.; Balaji, R.; Baldwin, E.; Barraclough, S.; Beane, C.; Bell, J.; Benford, T.; Bird, S.; Bishop, M.; Bloss, A.; Body, R.; Boulton, R.; Bown, A.; Bratten, C.; Bridgeman, C.; Britton, D.; Brooks, T.; Broughton-Smith, M.; Brown, P.; Buck, B.; Butcher, E.; Byrne, W.; Calderon, G.; Campbell, S.; Carr, O.; Carter, P.; Carter, D.; Cathrall, M.; Catton, M.; Chadwick, J.; Chapman, D.; Chau, K. K.; Chaudary, T.; Chidavaenzi, S.; Chilcott, S.; Choi, B.; Claasen, H.; Clark, S.; Clarke, R.; Clarke, D.; Clayton, R.; Collins, K.; Colston, R.; Connolly, J.; Cook, E.; Corcoran, M.; Corley, B.; Costello, L.; Coulson, C.; Crook, A.; Crook, D. W.; D'Arcangelo, S.; Darby, M. A.; Davis, J.; de Koning, R.; Derbyshire, P.; Devall, P.; Dolman, M.; Draper, N.; Driver, M.; Dyas, S.; Eaton, E.; Edwards, J.; Elderfield, R.; Ellis, K.; Ellis, G.; Elwell, S.; Evans, R.; Evans, B.; Evans, M.; Evans, R.; Eyre, D.; Fahey, C.; Fenech, V.; Field, J.; Field, A.; Foord, T.; Fowler, T.; French, M.; Fuchs, H.; Gan, J.; Gernon, J.; Ghadiali, G.; Ghuman, N.; Gibbons, K.; Gill, G.; Gilmour, K.; Goel, A.; Gordon, S.; Graham, T.; Grassam-Rowe, A.; Green, D.; Gronert, A.; Gumsley-Read, T.; Hall, C.; Hallis, B.; Hammond, S.; Hammond, P.; Hanney, B.; Hardy, V.; Harker, G.; Harris, A.; Havinden-Williams, M.; Hazell, E.; Henry, J.; Hicklin, K.; Hollier, K.; Holloway, B.; Hoosdally, S. J.; Hopkins, S.; Hughes, L.; Hurdowar, S.; Hurford, S. A.; Jackman, J.; Jackson, H.; Johns, R.; Johnston, S.; Jones, J.; Kanyowa, T.; Keating-Fedders, K.; Kempson, S.; Khan, I.; Khulusi, B.; Knight, T.; Krishna, A.; Lahert, P.; Lampshire, Z.; Lasserson, D.; Lee, K.; Lee, L. Y. W.; Legard, A.; Leggio, C.; Liu, J.; Lockett, T.; Logue, C.; Lucas, V.; Lumley, S. F.;

Maripuri, V.; Markham, D.; Marshall, E.; Matthews, P. C.; Mckee, S.; McKee, D. F.; McLeod, N.; McNulty, A.; Mellor, F.; Michel, R.; Mighiu, A.; Miller, J.; Mirza, Z.; Mistry, H.; Mitchell, J.; Moeser, M. E.; Moore, S.; Muthuswamy, A.; Myers, D.; Nanson, G.; Newbury, M.; Nicol, S.; Nuttall, H.; Nwanaforo, J. J.; Oliver, L.; Osbourne, W.; Osbourne, J.; Otter, A.; Owen, J.; Panchalingam, S.; Papoulidis, D.; Pavon, J. D.; Peace, A.; Pearson, K.; Peck, L.; Pegg, A.; Pegler, S.; Permain, H.; Perumal, P.; Peto, L.; Peto, T. E. A.; Pham, T.; Pickford, H. L.; Pinkerton, M.; Platton, M.; Price, A.; Protheroe, E.; Purnell, H.; Rawden, L.; Read, S.; Reynard, C.; Ridge, S.; Ritter, T. G.; Robinson, J.; Robinson, P.; Rodger, G.; Rowe, C.; Rowell, B.; Rowlands, A.; Sampson, S.; Saunders, K.; Sayers, R.; Sears, J.; Sedgewick, R.; Seeney, L.; Selassie, A.; Shail, L.; Shallcross, J.; Sheppard, L.; Sherkat, A.; Siddiqui, S.; Sienkiewicz, A.; Sinha, L.; Smith, J.; Smith, E.; Stanton, E.; Starkey, T.; Stawiarski, A.; Sterry, A.; Stevens, J.; Stockbridge, M.; Stoesser, N.; Sukumaran, A.; Sweed, A.; Tatar, S.; Thomas, H.; Tibbins, C.; Tiley, S.; Timmins, J.; Tomas-Smith, C.; Topping, O.; Turek, E.; Neibler, T.; Trigg-Hogarth, K.; Truelove, E.; Turnbull, C.; Tyrrell, D.; Vaughan, A.; Vertannes, J.; Vipond, R.; Wagstaff, L.; Waldron, J.; Walker, P.; Walker, A. S.; Walters, M.; Wang, J. Y.; Watson, E.; Webberley, K.; Webster, K.; Westland, G.; Wickens, I.; Willcocks, J.; Willis, H.; Wilson, S.; Wilson, B.; Woodhead, L.; Wright, D.; Xavier, B.; Yelnoorkar, F.; Zeidan, L.; Zinyama, R. COVID-19: Rapid Antigen Detection for SARS-CoV-2 by Lateral Flow Assay: A National Systematic Evaluation of Sensitivity and Specificity for Mass-Testing. *EClinicalMedicine* **2021**, *36*. https://doi.org/10.1016/j.eclinm.2021.100924.

(47) Stormonth-Darling, J. M.; Pedersen, R. H.; How, C.; Gadegaard, N. Injection Moulding of Ultra High Aspect Ratio Nanostructures Using Coated Polymer Tooling. *Journal of Micromechanics and Microengineering* **2014**, *24* (7), 075019. https://doi.org/10.1088/0960-1317/24/7/075019.

(48) Stormonth-Darling, J. M.; Pedersen, R. H.; Gadegaard, N. Polymer Replication Techniques. *Design of Polymeric Platforms for Selective Biorecognition* **2015**, 123–155. https://doi.org/10.1007/978-3-319-17061-9_6/FIGURES/9.

(49) Rihn, S. J.; Merits, A.; Bakshi, S.; Turnbull, M. L.; Wickenhagen, A.; Alexander, A. J. T.; Baillie, C.; Brennan, B.; Brown, F.; Brunker, K.; Bryden, S. R.; Burness, K. A.; Carmichael, S.; Cole, S. J.; Cowton, V. M.; Davies, P.; Davis, C.; de Lorenzo, G.; Donald, C. L.; Dorward, M.; Dunlop, J. I.; Elliott, M.; Fares, M.; da Silva Filipe, A.; Freitas, J. R.; Furnon, W.; Gestuveo, R. J.; Geyer, A.; Giesel, D.; Goldfarb, D. M.; Goodman, N.; Gunson, R.; James Hastie, C.; Herder, V.; Hughes, J.; Johnson, C.; Johnson, N.; Kohl, A.; Kerr, K.; Leech, H.; Lello, L. S.; Li, K.; Lieber, G.; Liu, X.; Lingala, R.; Loney, C.; Mair, D.; McElwee, M. J.; McFarlane, S.; Nichols, J.; Nomikou, K.; Orr, A.; Orton, R. J.; Palmarini, M.; Parr, Y. A.; Pinto, R. M.; Raggett, S.; Reid, E.; Robertson, D. L.; Royle, J.; Cameron-Ruiz, N.; Shepherd, J. G.; Smollett, K.; Stewart, D. G.; Stewart, M.; Sugrue, E.; Szemiel, A. M.; Taggart, A.; Thomson, E. C.; Tong, L.; Torrie, L. S.; Toth, R.; Varjak, M.; Wang, S.; Wilkinson, S. G.; Wyatt, P. G.; Zusinaite, E.; Alessi, D. R.; Patel, A. H.; Zaid, A.; Wilson, S. J.; Mahalingam, S. A Plasmid DNA-Launched SARS-CoV-2 Reverse Genetics System and Coronavirus Toolkit for COVID-19 Research. *PLoS Biol* **2021**, *19* (2), e3001091. https://doi.org/10.1371/JOURNAL.PBIO.3001091.

(50) Donald, C. L.; Brennan, B.; Cumberworth, S. L.; Rezelj, V. V.; Clark, J. J.; Cordeiro, M. T.; Freitas de Oliveira França, R.; Pena, L. J.; Wilkie, G. S.; Da Silva Filipe, A.; Davis, C.; Hughes, J.; Varjak,

M.; Selinger, M.; Zuvanov, L.; Owsianka, A. M.; Patel, A. H.; McLauchlan, J.; Lindenbach, B. D.; Fall, G.; Sall, A. A.; Biek, R.; Rehwinkel, J.; Schnettler, E.; Kohl, A. Full Genome Sequence and SfRNA Interferon Antagonist Activity of Zika Virus from Recife, Brazil. *PLoS Negl Trop Dis* **2016**, *10* (10). https://doi.org/10.1371/journal.pntd.0005048.

TOC

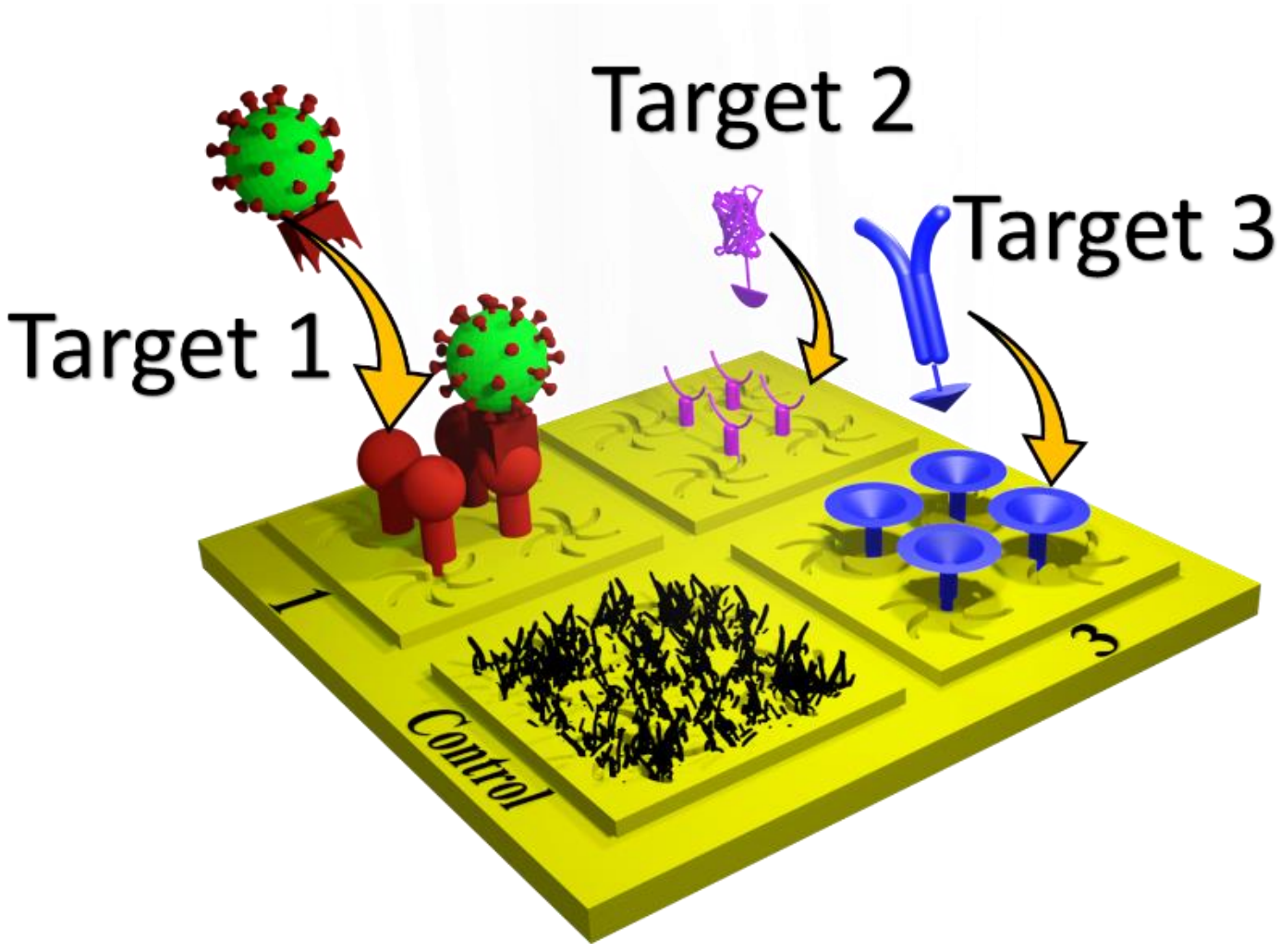